\documentclass[a4paper,12pt,epsf]{article}
\usepackage{epsfig} 
\input feynman
\newcommand{\newc}{\newcommand}
\newc{\lam}{\lambda}
\newc{\eps}{\epsilon}
\newc{\ra}{\rightarrow}
\newc{\wtilde}{\widetilde}
\begin{document}
\bigphotons
\setlength{\baselineskip}{.7cm}
%
%
%
%
%
\title{\textbf{Bounds on Broken $R$-Parity\\ from Leptonic Meson
    Decays}} \date{} \author{ \\Herbi
  Dreiner$^1$\footnote{dreiner@th.physik.uni-bonn.de},$~$ Giacomo
  Polesello$^2$\footnote{ polesell@hawk.pv.infn.it},$~$ Marc
  Thormeier$^3$\footnote{ thor@thphys.ox.ac.uk}\\
  \\
  \\
  $^1$ \emph{Physikalisches Institut der Universit\"at Bonn,}\\
  \emph{Nu\ss allee 12, 53115 Bonn, Germany}\\ \\
  $^2$ \emph{INFN, Sezione di Pavia,}\\\emph{Via Bassi 6, 27100 Pavia, 
Italy}\\ \\
  $^3$ \emph{Department of Theoretical Physics, University of
    Oxford,}\\\emph{1 Keble Road, Oxford OX1 3NP, United Kingdom}}
\maketitle
\begin{abstract}
  $~$\\
  Investigating leptonic decays of
  $\pi^-,K^-,B^-,\pi^0,K_L^0,B_s^0$ we present new bounds on some
  products of two $R$-parity violating coupling constants. For mesons
  of a similar structure but so far poor experimental data we give the
  corresponding formulae, to be used in the future.
\end{abstract}
%
%
%
%
%
\section{Introduction}
The $M\!S\!S\!M\!+$${\not\!\!R_p}$ is obtained from the $ M\! S\! S\!
M $ by adding the following terms to the superpotential ({\it c.f.}
ref.\cite{w})
\begin{eqnarray}\label{superpotential}
\Delta\mathcal{W}_{\not R_p}&=&\frac{1}{2}~\varepsilon^{ab}~\lam_{ijk}
~L^i_a~L^j_b~{E^k}^C~+~\varepsilon^{ab}~\delta^{xy}~
\lam^{\prime}_{ijk}~L^i_{a}~Q^j_{bx}~{D^{k}_y}^C~\nonumber\\
&~&~+~\frac{1}{2}~\varepsilon^{xyz}~\lam_{ijk}^{\prime\prime}
~{U^{i}_x}^C~{D^{{j}}_y}^C~{D^{k}_z}^C~+~\varepsilon^{ab}~\kappa_i
~L_a^i~H^{\mathcal{U}}_b.
\end{eqnarray}
$H,Q,L$ represent the leftchiral $SU\!(2)_W$-doublet superfields of
the Higgses, the quarks and leptons; $U,D,E$ represent the rightchiral
superfields of the $u$-type quarks, $d$-type quarks and electron-type
leptons, respectively; a superscript $^C$ denotes charge conjugation;
$a,b$ and $x,y,z$ are $SU\!(2)_W$- and $SU\!(3)_C$-indices, respectively;
 $~i,j,k$
and later also $f,g,l,n$ are generational indices (summation over
repeated indices is implied); $\delta^{xy}$ is the
Kronecker symbol, $\varepsilon^{...}$ symbolizes any tensor that is
totally antisymmetric with respect to the exchange of any two indices,
with $\varepsilon^{12...}=1$. The coupling constants
$\lam_{ijk}/\lam_{ijk}^{\prime\prime}$ are antisymmetric with
respect to the exchange of the first two/last two indices. The last
term in eq.(\ref{superpotential}) can be rotated away utilizing a
unitary field-redefinition.

Good agreement between $S\!M$ theory and experiment gives stringent
upper bounds on the extra 45 coupling constants $\lam_{ijk},
\lam_{ijk}^{\prime}$ and $\lam_{ijk}^{\prime \prime}$, as well as on
products thereof. For a list of references and the processes dealt
with, see e.g. ref.\cite{d,Bhattacharyya:1996nj,add}.  In particular,
in $\not\!\!R_p$ there are new operators for {\it leptonic} meson
decays. The $S\!M$ theoretical predictions for the decay widths of
mesons and the measured values match up within the experimental
uncertainty. We can thus determine  yet further tight
constraints on several products of coupling constants: $\lam^{\!  \prime
  \!*}\lam^ {\!\prime}$ and $\lam^{\!\prime\!*}\lam$.  This was first
done in ref.\cite{bgh} for single coupling constants and later in
ref.\cite{cdrs} for some products, however only treating charged pions
decaying via either $d$-type squark or slepton exchange, respectively.
Ref.\cite{bcd} treated general leptoquark reactions of several
particles, one of them the $K^0_L$; this result was quoted in terms of
$\not\!\!R_p$ by ref.\cite{elr}; the same result was reached by
ref.\cite{bsb}.  Ref.\cite{br} among other things dealt with the decay
of $K^0_L$, however with only $u$-squark exchange contributing to
$S\!M$-allowed processes. The decays of neutral and charged $B$-mesons
 were treated in ref.\cite{jkl} and ref.\cite{bk}, respectively. 
 We generalize these calculations, focusing
on products of two coupling constants, and stress where we
obtain new or stricter bounds.
\section{$\not\!\!R_{p}$-Decay of Charged Mesons}
\subsection{Calculation of the Decay Rate}
\begin{figure}
\begin{center}
\begin{picture}(32947,18099)
\THICKLINES
\put(0,0){${d^i}$}
\drawline\fermion[\NE\REG](1200,1200)[7000]
\drawarrow[\NE\ATTIP](3800,3800)
\put(\pbackx,\pbacky){$~\lam_{nik}^\prime P_L$}
\drawline\fermion[\SE\REG](\pbackx,\pbacky)[7000]
\drawarrow[\SE\ATTIP](9050,3250)
\put(\pbackx,0){${{\nu}^n}^C$}
\global\seglength=500
\global\gaplength=277
\drawline\scalar[\N\REG](\pfrontx,\pfronty)[9]
\drawarrow[\N\ATBASE](\pmidx,\pmidy)
\put(\pmidx,\pmidy){$~\!\!\!\!\!\!\!\!\!\!\!\!\wtilde{{d_R}^k}$}
\drawline\fermion[\NE\REG](\pbackx,\pbacky)[7000]
\drawarrow[\NE\ATBASE](8850,15550)
\put(\pfrontx,12300){$~\lam_{f\!jk}^{\prime*}P_R$}
\put(\pbackx,\pbacky){$~\ell^f$}
\drawline\fermion[\NW\REG](\pfrontx,\pfronty)[7000]
\drawarrow[\SE\ATTIP](\pmidx,\pmidy)
\put(\pbackx,\pbacky){$\!\!\!\!\!\!\!\!{u^j}^C$}
%
%
%
\put(15550,3500){$d^i$}
\drawline\fermion[\NE\REG](16750,4700)[7000]
\drawarrow[\NE\ATBASE](\pmidx,\pmidy)
\put(\pbackx,\pbacky)
{$\!\!\!\!\!\!\!\!\!\!\!\!\!\!\!\!\!\!\!\!\!\!\!\!\!\!\!-\lam_{kji}^
{\prime*}P_R~$}
\drawline\fermion[\NW\REG](\pbackx,\pbacky)[7000]
\drawarrow[\NW\ATBASE](\pmidx,\pmidy)
\put(\pbackx,\pbacky){$\!\!\!\!\!\!\!{u^j}$}
\global\seglength=500
\global\gaplength=277
\drawline\scalar[\E\REG](\pfrontx,\pfronty)[9]
\drawarrow[\E\ATBASE](\pmidx,\pmidy)
\put(\pmidx,7800){$\!\!\!\!\wtilde{{\ell_L}^k}$}
\drawline\fermion[\NE\REG](\pbackx,\pbacky)[7000]
\drawarrow[\NE\ATTIP](\pmidx,\pmidy)
\put(\pfrontx,9500){$~~\lam_{nk\!f}P_L$}
\put(\pbackx,\pbacky){$~\ell^f$}
\drawline\fermion[\SE\REG](\pfrontx,\pfronty)[7000]
\drawarrow[\NW\ATBASE](\pmidx,\pmidy)
\put(\pbackx,3500){$\nu^n$}
\end{picture}
\caption{\label{d}The tree-level $M\!S\!S\!M\!+$${\not\!\!R_p}$ processes 
contributing to the decay of the charged mesons.}
\end{center}
\end{figure}
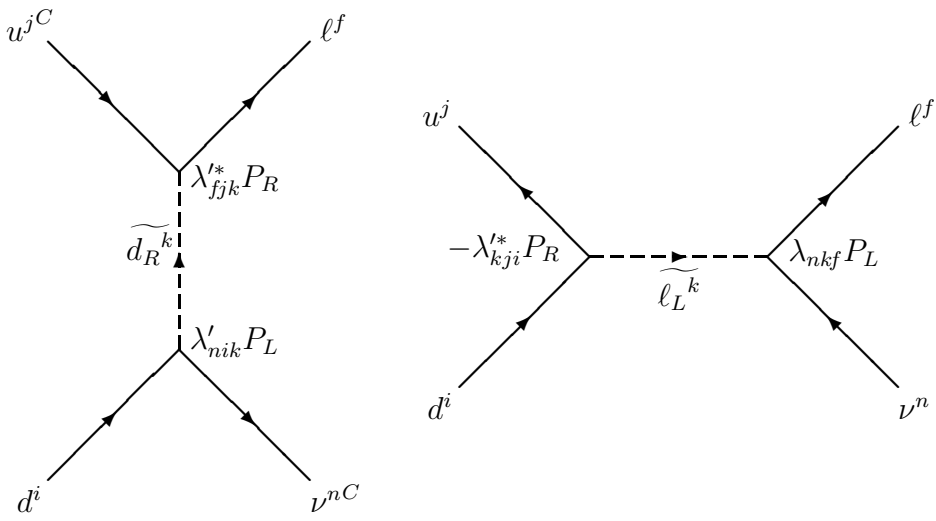

Consider a negatively charged meson $\pi^{ij}$ at rest made of a
$d$-type quark ${d}^i$ and a $u$-type antiquark ${{u}^j}^C$ which
decays into an antineutrino $ {{\nu}^n}^C$ and a charged lepton
$\ell^f$, i.e.
\begin{equation}\label{process}
\Big|~\pi^{ij}\!(p_1)~\Big\rangle~\longrightarrow~\Big|~{{\nu}^n}^C\!(p_2)~;
~\ell^f\!(p_3)~\Big\rangle,
\end{equation}
the $p_{1,2,3}$ being four-momenta.  We now calculate the partial
decay rate of this process. Focusing on the Yukawa-couplings of the
first two terms in eq.(\ref{superpotential}) leads to, again with
summation over repeated indices implied,
\begin{eqnarray}\label{lagrangian}
\mathcal{L}_{\not R_{p}}\supset&&\lam_{ijk}~\Big(~\overline{{{\nu}^{i}}^
{C}}~P_L~{\ell}^j~~\wtilde{{\ell_R}^{k}}^*~+~\overline{{\ell}^{k}}~P_L~
{\ell}^{j}~~\wtilde{{\nu_L}^{i}}+~\overline{{\ell}^{k}}~P_L~{\nu}^{i}~~
\wtilde{{\ell_L}^{j}}~\Big)\nonumber\\
&+&\lam^{\prime}_{ijk}~\Big(~\overline{{\nu^{i}}^{C}}~P_L~{d}^j~~
\wtilde{{d_R}^{k}}^{*}~+~\overline{{d}^{k}}~P_L~{d}^{j}~~
\wtilde{{\nu_L}^{i}}~+~\overline{{d}^{k}}~P_L~{\nu}^{i}~\wtilde{
{d_L}^{j}}~\Big)\nonumber\\
&-&\lam^{\prime}_{ijk}~\Big(~\overline{{{u}^{j}}^{C}}~P_L~{\ell}^i~~
\wtilde{{d_R}^{k}}^{*}~+~\overline{{d}^k}~P_L~{u}^{j}~\wtilde{
{\ell_L}^{i}}~+~\overline{{d}^{k}}~P_L~{\ell}^{i}~\wtilde{{u_L}^{j}}
~\Big)~~+~~c.c.~~~
\end{eqnarray}
All spinors are Dirac spinors, the overbar denotes the Dirac adjoint,
$P_{L,R}$ are the projection operators on the left-/right-handed
parts.  \emph{The fermions are mass-eigenstates}. A tilde denotes a scalar;
the scalars' subscripts $L,R$ indicate the chirality of the
corresponding Weyl spinor. The $4^{th}$ term in eq.(\ref{lagrangian})
together with the $c.c.$ of the $7^{th}$ term, and the $3^{rd}$ term
together with the $c.c.$ of the $8^{th}$ term lead to the meson decay
processes depicted in fig.  \ref{d}, which give the effective
Hamiltonians
\begin{eqnarray}\label{27}
\mathcal{H}^{\wtilde{d_R}}&=& \frac{1}{2}\sum_k\frac{{\lam}_{fjk}^
{\prime*}~\lam_{nik}^{\prime}}{m_{\wtilde{{d_R}^k}}^2}~~ \overline
{{\ell}^{f}}~\gamma_\nu~ P_L~\nu^{n}~~\overline{{u}^j}~\gamma^\nu~ P_L~d^i,
\nonumber\\
\mathcal{H}^{\wtilde{\ell_L}}&=&-~\sum_k\frac{{\lam}_{kji}^{\prime*} 
~\lam_{nkf}}{m_{\wtilde{{\ell_L}^k}}^2}~~\overline{{\ell}^{f}}~P_L~
{\nu}^n~~\overline{{u}^j}~P_R~{d}^i,
\end{eqnarray} 
\hspace{-0.28cm} where $m$ is the mass of a particle. To obtain the
first equation we employed a Fierz-identity. These two Hamiltonians
have to be added to the effective Hamiltonian for the $S\!M$-process
\begin{equation}\label{sm}
\mathcal{H}^W~=~\frac{4~G_F~V_{ji}}{\sqrt{2}}~~\overline{{\ell}^f}~
\gamma_{\nu}~P_L~\nu^f~~\overline{{u}^j}~\gamma^{\nu}~P_L~d^i.
\end{equation}
Here $G_F$ is the Fermi constant and $V_{ji}$ is an element of
the CKM-matrix. We obtain for the transition amplitude
$\mathcal{M}_{ijfn}$
\begin{equation}
\mathcal{M}_{ijfn}~\delta^4(p_1-p_2-p_3)~=~\frac{1}{2\pi i} \int \Big
\langle~\ell^f~;~{{\nu}^n}^C~\Big|~\Big(\mathcal{H}^{W} + \mathcal{H}^
{\wtilde{d_R}}+\mathcal{H}^{\wtilde{\ell_L}}\Big)~\Big|~\pi^{ij}~
\Big\rangle~d^4x.
\end{equation}
We expand the fields in the initial and final state, perform the
integrations and use
\begin{eqnarray}
\Big\langle~ 0~\Big|~\overline{u^j}\!(y)~\gamma^\nu~P_{L,R}~d^{i}\!(y)~\Big|~
\pi^{ij}\!(p_1)~\Big\rangle&=&\pm\frac{1}{\sqrt{2}}~f_{\pi^{ij}}~p^{\nu}_1~e^
{-ip_1y},\nonumber\\
\label{guglbaxi}
\Big\langle~ 0~\Big|~\overline{{u}^j}\!(y)~P_{L,R}~d^{i}\!(y)~\Big|~\pi^{ij}\!
(p_1)~\Big\rangle&=&\mp\frac{1}{\sqrt{2}}~\frac{m_{\pi^{ij}}^2}{m_{u^j}+m_{d^i
}}~f_{\pi^{ij}}~e^{-ip_1y};
\end{eqnarray}
$f_{\pi^{ij}}$ is the meson decay constant.\footnote{\label{footnote}There
 are several ways of defining the meson decay constant,
  differing by factors of $\sqrt{2}$; in the convention we use 
  $f_{\pi}= (92.4\pm 0.3)~$MeV, see ref.\cite{pdg}.}  
Thus
\begin{eqnarray}
\mathcal{M}_{ijfn}&=&~\frac{f_{\pi^{ij}}}{2\sqrt{2}}~\overline{\mathcal{U}^f}
\!(\vec{p}_3)~\times\nonumber\\
&~&\sum_k~\Bigg\{\Bigg(\frac{\delta_{fn}}{3}~\frac{8~G_F~V_{ji}}{\sqrt{2}}~+~
\frac{{\lam}_{fjk}^{\prime*}~{\lam}_{nik}^\prime}{m_{\wtilde{{d_R}^k}}^2}
\Bigg)\not\!p_1-2~\frac{m^2_{\pi^{ij}}}{m_{u^j}+m_{d^i}}~\frac{{\lam}_{kji}^{
\prime*}~\lam_{nkf}}{m^2_{\wtilde{{\ell_L}^k}}}\Bigg\}\nonumber\\
&\times&~~~P_L~\mathcal{V}^n\!(\vec{p}_2).
\end{eqnarray}
$\overline{\mathcal{U}^f},\mathcal{V}^n$ are the Fourier coefficient
functions of $\ell^f$, ${\nu^n}^C$, respectively.  Next we take the
absolute value square, average over the spins and use the trace
theorems. Then we sum over $n$, because the experiments that measured
the partial decay widths did not determine the flavour of the
antineutrinos,\footnote{The upper experimental bounds on $\pi^-\rightarrow
\mu{\nu_e}^C$ and $K^-\rightarrow\mu{\nu_e}^C$, see ref.\cite{pdg},
 come  from a different type of  experiment, compared to 
 the one used to determine  the branching ratios for $\pi^- 
\rightarrow\mu{\nu}^C$ and $K^- \rightarrow\mu{\nu}^C$. They do 
not lead to better bounds on the coupling constants.} resulting in
\begin{eqnarray}\label{mh2}
\sum_n \langle |\mathcal{M}_{ijfn}|^2\rangle=4G^2_Ff_{\pi^{ij}}^2|V_{ji}|
^2(m^2_{\pi^{ij}}-m^2_{\ell^f})~m^2_{\ell^f}\sum_n\Big(\delta_{fn}+2
\delta_{fn}\mbox{Re}[K_{ijfn}]+|K_{ijfn}|^2\Big),
\end{eqnarray}
with
\begin{equation}\label{NOMAD}
K_{ijfn}=\frac{\sqrt{2}}{8~G_F~|V_{ji}|}\sum_{k}\Bigg(~\frac{\lam^{\prime*}_
{fjk}~{\lam}^{\prime}_{nik}}{m_{\wtilde{{d_R}^k}}^2}-2~\frac{m^2_{\pi^{ij}}
}{m_{\ell^f}(m_{u^j}+m_{d^i})}~\frac{{\lam}^{\prime*}_{kji}~\lam_{nkf}}{m^2_{
\wtilde{{\ell_L}^k}}}\Bigg),
\end{equation}
containing all $\not\!\!R_p$ contributions; $2\mbox{Re}[K_{ijfn}]$ in
eq.(\ref {mh2}) is due to the interference between $S\!M$ and
$\not\!\!R_p$ amplitudes.  For simplicity we  neglect the phase
of the CKM-matrix. The partial decay rate is then
\begin{equation}\label{result}
\Gamma^{S\!M\!+\!{\not R_p}}_{\pi^{ij}\ra\ell^f\nu^C}~=~\Gamma_{\pi^
{ij}\ra\ell^f{\nu^f}^C}^{S\!M}~\times~\Big(1+2\mbox{Re}[K_{ijff}]+
\sum_n |K_{ijfn}|^2\Big),
\end{equation}
with $\nu^C$ being  an \emph{arbitrary} antineutrino, and   
\begin{equation}
\Gamma^{S\!M}_{\pi^{ij}\ra \ell^f {\nu^f}^C}=C_{ijf}~G_F^2~f_{\pi^
{ij}}^2~|V_{ji}|^2\frac{\Big((m_{\pi^{ij}}^2-m_{\ell^f}^2)~m_{\ell^f}\Big)^2}
{4\pi~ m_{\pi^{ij}}^3};
\end{equation}
the correction factor $C_{ijf}$ of $\mathcal{O}(1)$ is due to higher
order electroweak leading logarithms, short distance $QC\!D$
corrections, and structure dependent effects, see ref.\cite{ms} and
also ref.\cite{f}.
%
%
%
%
%
\subsection{Calculation of the Bounds}
We prefer not to compare the experimental data directly with
eq.(\ref{result}), since $f_{\pi^{ij}}$ has quite a large 
error. This leads to very weak bounds on $K_{ijfn}$. To avoid this,
we introduce
\begin{equation}\label{nar}
\mathcal{R}_{\pi^{ij}}:=\frac{\Gamma_{\pi^{ij}\ra \ell^f\nu^C}}{
\Gamma_{\pi^{ij}\ra \ell^{^g}\nu^C }},
\end{equation}
with $m_{\ell^{^g}}>m_{\ell^f}$. If the experimental and $S\!M
$-theoretical decay rates agree well we have, see eq.(\ref{result}),
$\Big|2\mbox{Re}[K_{ijff}]+\sum_n |K_{ijfn}|^2 \Big|\ll 1$. 
Plugging eq.(\ref{result})  into eq.(\ref{nar}) one gets 
\begin{eqnarray}
\frac{\mathcal{R}^{S\!M\!+\!{\not R_p}}_{\pi^{ij}}}{\mathcal{R}^
{S\!M}_{\pi^{
ij}}}:=1+\eps_{\pi^{ij}}\approx1+2~\mbox{Re}[K_{ijff}\!-\!K_{ijgg}]
+\sum_{n}|K_{ijfn}|^2-\sum_{n}|K_{ijgn}|^2.
\end{eqnarray}
Let $\Delta{...}$ symbolize the theoretical or experimental
uncertainty. If the theoretical prediction $\mathcal{R}^{S\!M}_{\pi
  ^{ij}}\!\pm\!\Delta\mathcal{R}^{S\!M}_{\pi^{ij}}$ lies within the
experimental range $\mathcal{R}^{exp}_{\pi^{ij}}\!\pm\!
\Delta\mathcal{R}^{exp}_{\pi^{ij}}$, one has
\begin{equation}\label{constraint}
\eps_{\pi^{ij}}^{min}:=\frac{\mathcal{R}^{exp}_{\pi^{ij}}} {\mathcal{R}^{S
\!M}_{\pi^{ij}}}-\Delta\Bigg(\frac{\mathcal{R}^{exp}_{\pi^{ij}}} {\mathcal{R}^
{S\!M}_{\pi^{ij}}}\Bigg)-1~\leq~\eps_{\pi^{ij}}~\leq~ \frac{\mathcal{R}^{
exp}_{\pi^{ij}}} {\mathcal{R}^{S\!M}_{\pi^{ij}}}+\Delta\Bigg(\frac{\mathcal{R}
^{exp}_{\pi^{ij}}} {\mathcal{R}^{S\!M}_{\pi^{ij}}}\Bigg)-1   =:\eps_{\pi^
{ij}}^{max}.
\end{equation} 
We could
use this to determine a bound on this general combination of 
$\not\!\!\!R_p$ coupling constants; however, the bounds on individual 
coupling constants  are typically of the order ${\cal O}(10^{-2})$,
see ref.\cite{d}, and thus we limit ourselves to at most two non-zero 
coupling constants  at a time, and in each case suppose the other 34 
$\lam,\,\lam'$ coupling constants vanish (eq.(\ref{tudel}),
eq.(\ref{result!!!}) and eq.(\ref{vfr})  are  also valid for  $f\rightarrow
 g$ with   $\epsilon^{max}_{\pi^{ij}}\leftrightarrow -
\epsilon^{min}_{\pi^{ij}}$):
\begin{eqnarray}\label{tudel}
\epsilon^{min}_{\pi^{ij}}~\leq~2~\mbox{Re}[K_{ijff}]~+~|K_{ijff}|^2
~\leq~\epsilon^{max}_{\pi^{ij}}~~~~\mbox{and for }n\neq f~~~~
|K_{ijfn}|^2~\leq~\epsilon^{max}_{\pi^{ij}}.
\end{eqnarray}
We assume that the imaginary parts of the coupling constants are 
 approximately the same as the corresponding real parts.\footnote{If
 the imaginary part vanishes the bounds are weaker by a factor 
of $\mathcal{O}(1)$.} With $G_F=(0.116639 \pm 0.000001)\times
(100~\mbox{GeV})^{-2}$, see
ref.\cite{pdg}, we obtain
\begin{eqnarray}\label{result0}\label{result!!!}
&&-0.330~|V_{ji}|~\Big(\sqrt{1+2~\eps_{\pi^{ij}}^{min}}+1\Big)~
\leq\nonumber\\
&~&~~~~~~~~~~~\frac{\mbox{Re}[{\lam}_{fjk}^{\prime*}~{\lam}_{fik}
^\prime]}{(m_{\wtilde{{d_R}^k}}/100~\mbox{GeV})^2}~,~~\frac{-2~
m_{\pi^{ij}}^2}{m_{\ell^f}(m_{u^j}+m_{d^i})}~\frac{\mbox{Re}[
{\lam}_{kji}^{\prime*}~\lam_
{fkf}]}{(m_{
\wtilde{{\ell_L}^k}}/100~\mbox{GeV})^2}\nonumber\\
&~&~~~~~~~~~~~~~~~~~~~~~~~\leq~0.330~|V_{ji}|~\Big(\sqrt{1+2
~\eps_{\pi^{ij}}^{max}}-1\Big)\end{eqnarray}
and for $n\neq f$ 
\begin{eqnarray}
\label{vfr}
 \frac{|\lam_{fjk}^{\prime*}~
\lam^{\prime}_{nik}|}{(m_{\wtilde{{d_R}^k}}/100~\mbox{GeV})^2}
~,~~\frac{2~m_
{\pi^{ij}}^2}{m_{\ell^f}(m_{u^j}+m_{d^i})}~\frac{|\lam_{kji}^
{\prime*}~
\lam_{nkf}|}{(m_{\wtilde{{\ell_L}^k}}/100~\mbox{GeV})^2}~\leq
~0.66~|V_{ji}|~\sqrt{\eps^{max}_{\pi^{ij}}}.
\end{eqnarray}
The prefactor $2m_{\pi^{ij}}^2/[m_{\ell^{f}}(m_{u^j}+m_{d^i})]$
 results in much
tighter bounds for $\lam_{kji}^{\prime*}~{\lam}_{nkf}$.  We will
 apply these results 
only to processes with sufficiently small experimental error bars.
%
%
%
%
%
\subsection{$\pi^-\ra\ell^{f,g}+\nu^C$}
As a first application, we consider pion decay with $f,i,j=1$, $g=2$.
The $S\!M$ gives the $2\sigma$ theoretical value $\mathcal{R}_{\pi^
  {-}}^{S\!M}=(1.2354 \pm 0.0004)\times10^{-4}$ (see ref.\cite{f};
the uncertainty mainly derives from $C_{111}$ and $C_{112}$). From the
partial decay widths at the $2\sigma$ level in ref.\cite{pdg}, namely
$\Gamma^{exp}_{\pi ^-\ra e\nu^C}$$/$$\Gamma^{exp}_{\pi^-~total}$$=$
$(1.230 \pm0.008)\times 10^{-4}~$ and $~\Gamma^{exp}_{\pi^- \ra \mu
  \nu^C}/\Gamma^{exp}_{\pi^-~total}=0.9998770\pm8\times10^{-7}$, one
calculates $\mathcal{R}_{\pi^-}^{exp}=(1.230\pm 0.008)~\times~10^{-4
  }$. Hence, $\eps^{min}_{\pi^{-}}=-0.0107$ and $\eps^{max}_{\pi^{-}
  }=0.0022$. With $|V_{11}|=0.9750\pm0.0008$, ref.\cite{bgh} obtained
bounds on a single coupling constant; this was updated in
ref.\cite{ls}. We have reproduced their results. The experimental data 
have only
marginally changed and the new bounds are   
$|\lam^{\prime}_{11k}| \leq
0.027~m_{{\wtilde{{d_R}^k}}}/100~\mbox{GeV}$ and $|\lam^{\prime}_
{21k}| \leq 0.059~m_{{\wtilde{{d_R}^k}}}/100~ \mbox{GeV}$. We
obtain bounds for the products of couplings $|\lam_{11k}^{\prime*}
~\lam^{\prime}_{21k}|\leq 0.03~(m_{\wtilde{{d_R}^k}}/100~\mbox{GeV})^2$,
 $|\lam_{11k}^{\prime*}~\lam^{\prime}_{31k}|\leq 0.03~(m_{\wtilde{{d_R}
^k}}/100~\mbox{GeV})^2$ and $|\lam_{21k}^{\prime*}~\lam^{\prime}_{31k}|
\leq 0.066~(m_{\wtilde{{d_R}^k}}/100~\mbox{GeV})^2$. The first bound is
 redundant since the product of the single bounds is
stronger; the second and the third bound are almost the same as the single
 bound on $|\lambda^\prime_{11k}|$ and $|\lambda^\prime_{21k}|$.
 Furthermore, we obtain the following new bounds using $m_e=
(0.510998902\pm2.1\times
10^{-8})~\mbox{MeV}$, $m_\mu=(105.6583568\pm5.2\times10^{-6})~\mbox
{MeV}$, $m_{\pi^-}=(139.57018\pm0.00035)~\mbox{MeV}$ and $m_u+m_{d}=
(8.5\pm3.5)~\mbox{MeV},$\footnote{This and $m_s$ are the biggest sources of
 inaccuracy, 
going linearly into the bounds on $\lam^{\prime*}\lam$.} see ref.\cite{pdg},
\begin{eqnarray}\label{keinenbockmehr}
-  7.9\times10^{-8}~\Bigg(\frac{m_{\wtilde{{\ell_L}^k}}}{100~\mbox{GeV}}
\Bigg)^2 \leq &\mbox{Re}[\lam^{\prime*}_{k11}~\lam_{1k1}]&\leq 7.1\times10^
{-5}~\Bigg(\frac{m_{\wtilde{{{\ell_L}}^k}}}{100~\mbox{GeV}}\Bigg)^2 
           ,\nonumber\\
-   7.9\times10^{-5}~\Bigg(\frac{m_{\wtilde{{\ell_L}^k}}}{100~\mbox{GeV}}
\Bigg)^2   \leq&  ~\mbox{Re}[\lam^{\prime*}_{311}~\lam_{232}]&, 
\nonumber\\
|\lam^{\prime*}_{k11}~\lam_{3k1}| &\leq&3.4\times10^{-6}\Bigg(\frac{m_{
\wtilde{{\ell_L}^k}}}{100~\mbox{GeV}}\Bigg)^2,           \nonumber\\
|\lam^{\prime*}_{211}~\lam_{322}| &\leq&1.5\times10^{-3}\Bigg(\frac{m_{
\wtilde{{\ell_L}^k}}}{100~\mbox{GeV}}\Bigg)^2,\nonumber\\
|\lam^{\prime*}_{111}~\lam_{211}| &\leq&3.4\times10^{-6}\Bigg(\frac{m_{
\wtilde{{\ell_L}^k}}}{100~\mbox{GeV}}\Bigg)^2.       
\end{eqnarray}
The upper bound/bounds we obtained for $\mbox{Re}[\lam^{\prime*}_{311}
~\lam_{232}]$/$\mbox{Re}[\lam^{\prime*}_{111}~\lam_{212}]$, $|
\lambda^{\prime*}_{111}~\lambda_{312}|$ are weaker
than the products of the two bounds on the single coupling constants, see
 ref.\cite{add}; there also much stricter bounds were stated for 
$|\lam^{\prime*}_{k11}~\lam_{1k2}|$ as well as for $|\lambda^{\prime*}_{311}~
\lambda_{231}|$.
%
%
%
%
%
\subsection{$K^- \ra \ell^{f,g} + \nu^C$}
Next we consider charged kaon decay with $f,j=1,~g,i=2$. According to
ref.\cite{f}, $\mathcal{R}_{K^-}^{S\!M} =(2.472\pm0.002)\times10^{-5}$
at the $2\sigma$ level. Experimentally $\Gamma^{exp}_{K^-\ra
  e\nu^C}$$/$$\Gamma^{exp}_{K^- ~total} =(1.55\pm0.14)\times10^{-5}~$
and $~\Gamma^{exp}_{K^-\ra\mu\nu^C}$$/$$\Gamma^{exp}_{K^-~total}=
0.6351\pm0.0036$, at the $2\sigma$ level \cite{pdg}. Therefore
$\mathcal{R}_{K^-}^{exp}=(2.44\pm 0.22)~\times~10^{-5}$ and $\eps
^{min}_{K^{-}}=-0.10$ and $\eps^{max}_{K^{-}}=0.076$. Using $|V_{12
  }|=0.222\pm0.004$ we obtain $|\lam_{11k}^{\prime*}~\lam^{\prime}_{32k}|
\leq 0.04~(m_{\wtilde{{d_R}^k}}/100~\mbox{GeV})^2$, $|\lam_{21k}^{\prime*}
~\lam^{\prime}_{32k}|\leq0.046~(m_{\wtilde{{d_R}^k}}/100~\mbox{GeV})^2$.
As for the pion these bounds are almost the same as the
 ones on $|\lambda^\prime_{11k}|$, $|\lambda^\prime_{21k}|$.  
Our bounds on $\mbox{Re}[\lam^{\prime*}_{11k}~\lam^{\prime}_{12k}],~
\mbox{Re}[\lam^{\prime*}_{21k}~\lam^{\prime}_{22k}]$ are much weaker
than the bounds on $|\lam^{\prime*}_{i1k}~\lam^\prime_{i2k}|$, see 
ref.\cite{add}, and we do not list them. Similarly, the
existing bounds on $ |\lam_{11k}^{\prime*}~\lam^{\prime}_{22k}|,~|
\lam_{21k}^{\prime*}~\lam^{\prime}_{12k}| $ are much stronger than
ours.  Furthermore with $m_{K^-}=(493.677\pm0.016)~\mbox{MeV}$, $m
_s=(122.5\pm47.5)~\mbox{MeV}$ and $m_s=(21\pm4)m_d$ (see
ref.\cite{pdg}) we have the following new bounds
\begin{eqnarray}
-  7.0\times10^{-7}~\Bigg(\frac{m_{\wtilde{{\ell_L}^k}}}{100~\mbox{GeV}}
\Bigg)^2\leq&  \mbox{Re}[\lam^{\prime*}_{k12}~\lam_{1k1}]&\leq 1.8
\times10^{-5}~\Bigg(\frac{m_{\wtilde{{\ell_L}^k}}}{100~\mbox{GeV}}\Bigg)^2
 ,\nonumber\\
-1.8\times10^{-4}~\Bigg(\frac{m_{\wtilde{{\ell_L}^k}}}{100~\mbox{GeV}}
\Bigg)^2  \leq&  \mbox{Re}[\lam^{\prime*}_{k12}~\lam_{2k2}]&\leq 3.8
\times10^{-3}~\Bigg(\frac{m_{\wtilde{{\ell_L}^k}}}{100~\mbox{GeV}}
\Bigg)^2~(k=3),\nonumber\\
|\lam^{\prime*}_{k12}~\lam_{2k1}| &\leq&5.4\times10^{-6}\Bigg(\frac{m_{
\wtilde{{\ell_L}^k}}}{100~\mbox{GeV}}\Bigg)^2,          \nonumber\\
|\lam^{\prime*}_{k12}~\lam_{3k1}| &\leq&5.4\times10^{-6}\Bigg(\frac{m_{
\wtilde{{\ell_L}^k}}}{100~\mbox{GeV}}\Bigg)^2,           \nonumber\\
|\lam^{\prime*}_{k12}~\lam_{1k2}| &\leq&1.3\times10^{-3}\Bigg(\frac{m_{
\wtilde{{\ell_L}^k}}}{100~\mbox{GeV}}\Bigg)^2,           \nonumber\\
|\lam^{\prime*}_{k12}~\lam_{3k2}| &\leq&1.3\times10^{-3}\Bigg(\frac{m_{
\wtilde{{\ell_L}^k}}}{100~\mbox{GeV}}\Bigg)^2 .
\end{eqnarray}
The upper bound on $\mbox{Re}[\lam^{\prime*}_{k12}~\lam_{212}]$ obtained from 
 two bounds on the single coupling constants is stricter than the 
one we obtained.  
%
%
%
%
\subsection{$B^-\ra\ell^f+\nu^C$}

For the charged $B$-meson decay the procedure is slightly different
since it has not been directly measured. Unlike the two previous cases
one only has an experimental upper bound on the branching ratio $\mathcal{B}$,
 see
ref.\cite{pdg}, and  thus has to go back to eq.(\ref{result}).
This has been done in ref.\cite{bk}. We go beyond their 
work with a more conservative account of the experimental errors and
obtain weaker bounds. We also work from the beginning in the mass
eigenstate basis to avoid model dependent results, see ref.\cite{dm}.

First $f=3$. The theoretical predictions are limited by
$\Gamma^{S\!M\!+\!{\not R_p}}_{B\ra\tau\nu^C}/\Gamma^{S\!M\!+\!{\not
R_p}}_{B~{total}}$ $\leq~5.7\times10^{-4}$. As the total widths 
$\Gamma^{exp}_{B~total}$ and $\Gamma^{S\!M}_
{B~total}$ agree fairly well one has $\Gamma^{S\!M\!+\!{\not
    R_p}}_{B~total}\approx\Gamma^{S\!M}_{B~total}$, so that, utilizing
eq.(\ref{result}), we obtain for the branching ratio
\begin{equation}\label{34}
\frac{\Gamma^{S\!M\!+\!{\not R_p}}_{B\ra\tau\nu^C}}{\Gamma^{S\!M\!+\!{
\not R_p}}_{B~total}}~\approx~\Big(1+2~\mbox{Re}[K_{3133}]+\sum_n|K_{313n}|^2
\Big)~\frac{\Gamma^{S\!M}_{B\ra\tau{\nu_\tau}^C}}{\Gamma^{S\!M}_{B~total}}.
\end{equation}
To keep the combined  uncertainties of $|V_{13}|$ and $f_B$ as small as
possible we use the theoretical prediction (see ref.\cite{cleo})
\begin{equation}\label{35}
\frac{\Gamma^{S\!M}_{B\ra\tau{\nu_\tau}^C}}{\Gamma^{S\!M}_{B~total}}~=
~\Big(4.08
\pm0.24\Big)~\times~10^{-4}~\Bigg|\frac{V_{13}}{V_{31}}\Bigg|^2.
\end{equation}
In order to take into account the correlated uncertainties in $V_{13}/V_{31}$
we use the Wolfenstein parameterization  (see e.g. ref.\cite{b}): 
\begin{eqnarray}\label{kobel}
\frac{V_{13}}{V_{31}}=\frac{\bar{\rho}-i\bar{\eta}}{1-\frac{\lam^2}{2}-\bar{
\rho}-i\bar{\eta}}.
\end{eqnarray}
The Wolfenstein parameters are given by (see ref.\cite{n}) $\bar{
  \rho}=0.21\pm0.12,~\bar{\eta}=0.38\pm 0.11,~\lam=0.222\pm0.004$, all
at 95\% C.L.. We thus obtain for the theoretical prediction $\Gamma^
{S\!M}_ {B\ra\tau{\nu_\tau}^C}/\Gamma^{S\!M}_{B~total}=(1.05\pm0.65)~\times
~10^{-4}$. The lower value should be used in eq.(\ref{34}), to be
compared with the experimental upper bound. Thus
\begin{equation}
2~\mbox{Re}[K_{3133}]+\sum_n|K_{313n}|^2~\leq~13.3.
\end{equation}
In the following, 
we again assume that only two coupling
constants are non-zero.  Thus we have
$|K_{3131}|,|K_{3132}|\leq\sqrt{13.3}$.  Furthermore  the 
 imaginary part is again taken to be  about  the same as the real part,
hence $-\sqrt{\frac{1}
  {4}+\frac{13.3}{2}}-\frac{1}{2}\leq\mbox{Re}[K_{3133}]\leq\sqrt{\frac{1}
  {4}+\frac{13.3}{2}}-\frac{1}{2}$.  Thus, with $|V_{13}|=0.0035\pm0.0015$
 (see
ref.\cite{pdg}), $m_{B^-}=(5279.0\pm0.5)~\mbox{MeV}$, $m_b=(4200\pm200)
~\mbox{MeV}$ and $m_\tau=(1776.99\pm0.29)~\mbox{MeV}$, see ref.\cite{pdg}
, we obtain
\begin{eqnarray}
|\lam^{\prime*}_{313}~\lam_{233}|&\leq&2\times10^{-3}~\Bigg(\frac{m_{
\wtilde{{\ell_{L}}^k}}}{100~\mbox{GeV}}\Bigg)^2,\nonumber\\
-6\times10^{-4}~\Bigg(\frac{m_{\wtilde{{\ell_{L}}^k}}}{100~\mbox{GeV}}
\Bigg)^2~\leq~\mbox{Re}[\lam^{\prime*}_{213}~\lam_{323}]&\leq&1\times10^{-3}~
\Bigg(\frac{m_{\wtilde{{\ell_{L}}^k}}}{100~\mbox{GeV}}\Bigg)^2.
\end{eqnarray}
According to ref.\cite{bk} the bounds on $|\lam^{\prime*}_{k13}~
\lam_{1k3}|$, $|\lam^{\prime*}_{31k}~\lam^\prime_{13k}|$, $|\lam
^{\prime*}_{31k}~\lam^\prime_{23k}|$, $\mbox{Re}[\lam^{\prime*}
_{31k}~\lam^\prime_{33k}]$ and $\mbox{Re}[\lam^{\prime*}_{113}
~\lam_{313}]$  are not better than the previous ones; furthermore
 the bound on $|\lambda^{\prime*}_{113}~\lambda_{213}|$ is weaker than
the product of the two bounds on the single coupling constants, see
ref.\cite{add}.

Analogously, for $f=1,2$,
\begin{equation}
\Big(1+2~\mbox{Re}[K_{31ff}]+\sum_n|K_
{31fn}|^2\Big)~\Gamma^{S\!M}_{B\ra\ell^f{\nu^f}^C}\times \tau_{B^-}<
5.7~\times~10^{-4},
\end{equation}
where $\tau_{B^-}$ is the $B$-meson life time. Instead of arguing that the
error on $\Gamma^{S\!M}_{B\ra\ell^f{\nu^f}^C}$ is
$\pm\frac{0.65}{1.05} \Gamma^ {S\!M}_{B\ra\ell^f{\nu^f}^C}$, see the third
 line below eq.(\ref{kobel}), we are
going to be as conservative as possible. Due to isospin invariance
$B^0$ and $B^-$ have the same decay constant. From ref.\cite{bern},
$f_{B^0}=(200\pm 30)$ MeV, and thus with our convention $f_B=(141\pm
21)$ MeV ({\it c.f.} footnote \ref{footnote}). Therefore, $f_B^2\,
|V_{13}|^2=(0.24\pm 0.22)\,$MeV$ ^2$, and with $\tau_{B^-}=1.655\times
10^{-12~}s$ (see ref.\cite{pdg}), we obtain for $f=1$ that $(1+...)
(9.0\pm8.3)\times 10^{-12}\leq1.5\times 10^{-5}$ and for $f=2$ that
$(1+...)(3.8 \pm3.5 )\times10^{-7}\leq2.1 \times10^{-5}$.  Working
with the lower value, for $f=1$ we get
\begin{eqnarray}
|\lam^{\prime*}_{k13}~\lam_{3k1}|&\leq&6\times10^{-4}~\Bigg(\frac{m_{
\wtilde{{\ell_{L}}^k}}}{100~\mbox{GeV}}\Bigg)^2.
\end{eqnarray} 
The bounds on $|\lam^{\prime*}_{k13}~\lam_{2k1}|$ and  $\mbox{Re}[\lam
^{\prime*}_{k13}~\lam_{1k1}]$ are not improved compared to the 
previous ones, see ref.\cite{bk}, and the ones  on $|\lam^{\prime*}_{11k}~\lam
^{\prime}_{23k}|$, $|\lam^
{\prime*}_{11k}~\lam^{\prime}_{33k}|$ and Re$[\lam^{\prime*}
_{11k}~\lam^{\prime}_{13k}]$ are too poor to be listed.  $f=2$
yields
\begin{eqnarray}
|\lam_{k13}^{\prime*}\lam_{3k2}|&\leq&7\times 10^{-4}~\Bigg(
\frac{m_{\widetilde{{\ell_L}^k}}}{100~\mbox{GeV}}\Bigg)^2.
\end{eqnarray} 
Ref.\cite{bk} states that there exist better bounds on 
$|\lam^{\prime*}_{k13}~\lam_{1k2}|$ and $\mbox{Re}
[\lam^{\prime*}_{k13}~\lam_{2k2}]$. The bounds on 
$|\lam^{\prime*}_{21k}~\lam^\prime_{13k}|,|\lam^{
  \prime*}_{21k}~\lam^\prime_{33k}|,\mbox{Re}[\lam^{\prime*}_
{21k}~\lam^\prime_{23k}]$ are almost the same as
the single bound on $|\lam^{\prime}_{21k}|$.
%
%
%
%
%
%
%
\section{$\not\!\!R_{p}$-Decay of  Neutral Mesons}
\subsection{\label{bounds}Calculation of the Bounds}
Now we deal with the bound state $({d^j}^C\!d^i)$ decaying into $\ell
^f$ and ${\ell^n}^C$, with momenta $p_1,p_3,p_2$, respectively. We
consider only $n\!\neq\!f$, in which case the process does not occur
in the $S\!M$ and therefore no contributions from loop-diagrams have to
be taken into account.  We proceed as in the previous section. In
eq.(\ref{lagrangian}) the $9^{th}$ term together with its $c.c.$
contributes to the decay, in analogy to the $\wtilde{{d_R}^k}
$-exchange in the last section. Furthermore the $2^{nd}$ term together
with the $c.c.$ of the $5^{th}$ and the $5^{th}$ term together with
the $c.c.$ of the $2^{nd}$ contribute, both in analogy to the
$\wtilde{{\ell_L}^k}$-exchange. The Hamiltonian is given by
\begin{eqnarray}
\mathcal{H}^{\wtilde{u_L}}&=\frac{1}{2}&\sum_k\frac{\lam_{nkj}^{\prime}~
\lam_{f\!ki}^{\prime*}}{m^2_{\wtilde{{u_L}^k}}}~~\overline{\ell^f}~\gamma^
\nu P_L~\ell^n~~\overline{d^j}~\gamma_\nu P_R~d^i,\nonumber\\
\mathcal{H}^{\wtilde{\nu_L}}&=&\sum_k\frac{\lam_{kji}^{\prime*}~\lam_{kn\!f}}
{m^2_{\wtilde{{\nu_L}^k}}}~~\overline{\ell^f}~P_L~\ell^n~~\overline{d^j}~
P_R~d^i\nonumber\\
&+&\sum_k\frac{\lam_{kij}^{\prime}~\lam_{k\!fn}^{*}}{m^2_{\wtilde{{\nu_L}^k}}}
~~\overline{\ell^f}~P_R~\ell^n~~\overline{d^j}~P_L~d^i.
\end{eqnarray}
Using the results corresponding to eq.(\ref{guglbaxi}) we obtain
\begin{eqnarray}\label{ts}
\mathcal{M}_{ijfn}=-\frac{f_{({d^j}^C\!d^i)}}{2\sqrt{2}}~\overline{
\mathcal{U}^f}\!(\vec{p}_3)~\Big\{~A_{ijfn}~\not\!p_1~P_L~+~B_{ijfn}~
P_L~-~B_{jinf}^*~P_R~\Big\}~\mathcal{V}^n(\vec{p}_2),
\end{eqnarray}
where
\begin{eqnarray}
A_{ijfn}=\sum_k\frac{\lam^{\prime}_{nkj}~\lam^{\prime*}_{fki}}{m_{
\wtilde{{u_L}^k}}^2}~=~A_{jinf}^*,\qquad B_{ijfn}&=&2\frac{m_{({d^j}^
C\!d^i)}^2}{(m_{d^j}+m_{d^i})}~\sum_k\frac{\lam^{\prime*}_{kji}~
\lam_{kn\!f}}{m_{\wtilde{{\nu_L}^k}}^2}.~~~~  
\end{eqnarray}
Hence 
\begin{eqnarray}
\Gamma^{S\!M\!+\!{\not R_p}}_{({d^j}^C\!d^i)\ra\ell^f+{\ell^n}^C} &=&\sqrt
{m_{({d^j}^C\!d^i)}^4\!+m_{\ell^f}^4+m_{\ell^n}^4 - 2 \Big(m_{({d^j}^C\!d^i)}
^2 m_{\ell^f}^2+m_{\ell^f}^2m_{\ell^n}^2+m_{\ell^n}^2m_{({d^j}^C\!d^i)}^2
\Big)~~ }\nonumber\\
& \times & \frac{f_{({d^j}^C\!d^i)}^2}{128~\pi~m_{({d^j}^C\!d^i)}^3}~\times~
\Upsilon_{ijfn}, 
\end{eqnarray}
where
\begin{eqnarray}
\Upsilon_{ijfn}&=&\Big(m_{({d^j}^C\!d^i)}^2-m_{\ell^f}^2\Big)\Big|A_{ijfn}~
m_{\ell^f}+B_{ijfn}\Big|^2+\Big(m_{({d^j}^C\!d^i)}^2-m_{\ell^n}^2\Big)\Big
|A_{ijfn}~m_{\ell^n}+B_{jinf}^*\Big|^2\nonumber\\
&- &\Big|B_{ijfn}~m_{\ell^n}-B_{jinf}^*~m_{\ell^f}\Big|^2~+~m_{\ell^f}~
m_{\ell^n}\Bigg\{\Big|B_{ijfn}+B_{jinf}^*\Big|^2~-~\Big|A_{ijfn}~m_{\ell^n}
-B_{ijfn}\Big|^2\nonumber\\
&-&\Big|A_{ijfn}~m_{\ell^f}-B_{jinf}^*\Big|^2~+~\Big|(m_{\ell^f}+m_{\ell^n})
~A_{ijfn}~\Big|^2\Bigg\}.
\end{eqnarray}
Due to the large experimental error in $f_{({d^j}^C\!d^i)}$, we can
neglect $m_{\ell^n}$ compared to $m_{\ell^f}$ (with $f,n$ chosen
correspondingly) and $m_{\ell^f}$ compared to $m_{({d^j}^C\!d^i)}$.
Thus, focusing again on the bounds on products of two coupling
constants, with all other coupling constants vanishing,
\begin{equation}\label{r3}
|A_{ijfn}|,\frac{|B_{ijfn}|}{m_{\ell^f}},\frac{|B_{jinf}^*|}{m_{\ell^f}}~\leq~
\frac{20}{f_{({d^j}^C\!d^i)}~m_{\ell^f}}~ \sqrt{\frac{\Gamma^
{exp.~upper~bound}_{({d^j}^C\!d^i)\ra\ell^f+{\ell^n}^C}\Big/\Gamma^{exp}_
{({d^j}^C\!d^i)\ra total}}{m_{({d^j}^C\!d^i)}~\tau_{({d^j}^C\!d^i)}^{exp}}~}.
\end{equation}
Here $\tau$ is the mean life time. The same considerations apply to
mesons that have wave functions of the form
\begin{equation}
\pi_{ij}^0=\frac{1}{\sqrt{2}}\Big[ ({d^j}^C\!d^i)\pm({d^i}^C\!d^j)\Big];
\end{equation}
one replaces every $A_{ijfn}$ by $\frac{1}{\sqrt{2}}(A_{ijfn}\pm
A_{jifn})$, and likewise for $B_{ijfn}$, $B^*_{jinf}$. As in the
previous section, we will apply eq.(\ref{r3}) only to processes with
satisfactory experimental data, as was done in ref.\cite{jkl}, treating
among other processes  $B^0\rightarrow\ell^f+{\ell^n}^C$; we confirm 
their results.

\subsection{$B^0_s\ra\mu+e^C$}
We now consider $B^0_s,{B^0_s}^C\ra\mu+e^C$, $i,f=2,j=3,n=1$ and 
$i\leftrightarrow j$.
 The relevant parameters are given by $f_{B^0_s}=(1.16\pm0.04)~f_{B^0}$,
see ref.\cite{bern}, ${\cal B}(B^0_s\ra\mu+e^C)<6.1\times10^{-6}$, 
$\tau_{B^0_s}=(1.464 \pm0.057)\times10^{-12}~s$, $m_{B^0_s}=(5369.6
\pm2.4)~$MeV, see ref.\cite{pdg}.
Thus
\begin{eqnarray}
|\lam^\prime_{123}~\lam^{\prime*}_{222}|&\leq&8\times10^{-3}\Bigg(\frac{m_{
\wtilde{{u_L}^k}}}{100~\mbox{GeV}}\Bigg)^2,\nonumber\\
|\lam^\prime_{1k2}~\lam^{\prime*}_{2k3}|&\leq&8\times10^{-3}\Bigg(\frac{m_{
\wtilde{{u_L}^k}}}{100~\mbox{GeV}}\Bigg)^2,\nonumber\\
|\lam^{\prime*}_{k32}~\lam_{k12}|&\leq&7\times10^{-5}\Bigg(\frac{m_{\wtilde{{
\nu_L}^k}}}{100~\mbox{GeV}}\Bigg)^2,\nonumber\\
|\lam^{\prime*}_{k23}~\lam_{k21}|&\leq&7\times10^{-5}\Bigg(\frac{m_{\wtilde{{
\nu_L}^k}}}{100~\mbox{GeV}}\Bigg)^2,\nonumber\\
|\lam^{\prime*}_{k32}~\lam_{k21}|&\leq&7\times10^{-5}\Bigg(\frac{m_{\wtilde{{
\nu_L}^k}}}{100~\mbox{GeV}}\Bigg)^2,\nonumber\\
|\lam^{\prime*}_{k23}~\lam_{k12}|&\leq&7\times10^{-5}\Bigg(\frac{m_{\wtilde{{
\nu_L}^k}}}{100~\mbox{GeV}}\Bigg)^2.
\end{eqnarray} 
Our results for $|\lam^\prime_{1k3}~\lam^{\prime*}_{2k2}|$ ($k\neq2$)
 and $k=1$ for the  second bound are weaker than the products of the bounds on
 single coupling constants, see ref.\cite{add}.  
\subsection{$K^0_{L}\ra\mu+e^C$}
$K^0_{L}$ is defined as $[K_2^0+\eps K_1^0]/\sqrt{1+\eps^2}$, with
$K^0_{1,2}=[K^0 \pm {K^0}^C]/\sqrt{2}$. $\eps$ parameterizes the
$C\!P$-violation. If we neglect $\eps$, $K^0_L=[K^0-{K^0}^C]/\sqrt
{2}$, with $K^0=(s^Cd)$. From ref.\cite{pdg} one has $m_{K^0_L}=
(497.672\pm 0.031)~$MeV, $\tau_{K^0_L}=(5.17\pm0.04)\times10^{-8}~s$
and ${\cal B}(K^0_{L}\ra\mu+e^C)<4.7\times10^{-12}$. Ref.\cite{pdg}
gives $f_{K}=(159\pm1.4\pm0.44)~$MeV, which in the convention we use
gives the central value $112.4~$MeV. Hence, the first two bounds
updating previous ones,
\begin{eqnarray}
|\lam^\prime_{1k2}~\lam^{\prime*}_{2k1}|&\leq&3\times10^{-7}\Bigg(\frac{m_{
\wtilde{{u_L}^k}}}{100~\mbox{GeV}}\Bigg)^2,\nonumber\\
|\lam^\prime_{1k1}~\lam^{\prime*}_{2k2}|&\leq&3\times10^{-7}\Bigg(\frac{m_{
\wtilde{{u_L}^k}}}{100~\mbox{GeV}}\Bigg)^2,\nonumber\\
|\lam^{\prime*}_{k21}~\lam_{k12}|&\leq&6\times10^{-9}\Bigg(\frac{m_{\wtilde{{
\nu_L}^k}}}{100~\mbox{GeV}}\Bigg)^2,\nonumber\\
|\lam^{\prime*}_{k12}~\lam_{k12}|&\leq&6\times10^{-9}\Bigg(\frac{m_{\wtilde{{
\nu_L}^k}}}{100~\mbox{GeV}}\Bigg)^2,\nonumber\\
|\lam^{\prime}_{k12}~\lam_{k21}^*|&\leq&6\times10^{-9}\Bigg(\frac{m_{\wtilde{{
\nu_L}^k}}}{100~\mbox{GeV}}\Bigg)^2,\nonumber\\
|\lam^{\prime}_{k21}~\lam_{k21}^*|&\leq&6\times10^{-9}\Bigg(\frac{m_{\wtilde{{
\nu_L}^k}}}{100~\mbox{GeV}}\Bigg)^2.
\end{eqnarray} 
\subsection{$\pi^0\ra\mu+e^C$}
With small modifications the result of section \ref{bounds} can also
be carried over to admixtures of $({d^j}^C\!d^i)$ with
$({u^j}^C\!u^i)$, as the latter term does not contribute to any decay
because the $u$-type quarks do not couple together to the $\not\!\!\!R_p~$
operators. However, we shall limit ourselves to the $\pi^0$:
$\eta$ and $\eta'$ are more complicated, see ref.\cite{pdg}, and the
experimental data do not suffice to extract satisfactory bounds.
 
The relevant parameters here are $m_{\pi^0}=(134.9766\pm0.0006)~$MeV,
$\tau_{\pi^0}=(8.4\pm0.6)\times10^{-17}~s$ and ${\cal B}(\pi^0
\ra\mu+e^C)<3.8\times10^{-10}$, see ref.\cite{pdg}. Thus
\begin{eqnarray}
|\lam^{\prime*}_{311}~\lam_{312}|&\leq&3\times10^{-3}\Bigg(\frac{m_{\wtilde{{
\nu_L}^k}}}{100~\mbox{GeV}}\Bigg)^2,\nonumber\\
|\lam^{\prime}_{311}~\lam_{321}^*|&\leq&3\times10^{-3}\Bigg(\frac{m_{\wtilde{{
\nu_L}^k}}}{100~\mbox{GeV}}\Bigg)^2.
\end{eqnarray}
In ref.\cite{add} a much stronger bound is stated for $|\lam^\prime
_{1k1}~\lam^{\prime*} _{2k1}|$. Furthermore they present a better bound
 for $|\lambda_{111}^{\prime}~\lambda_{121}^*|$; and from ref.\cite{add}
one finds a stricter bound on $|\lam^{\prime*}_{211}~\lam_{212}|$, based
on the bounds on single coupling constants.
%
%
\section{Summary} 
We have determined the bounds on products of $\not\!\!\!R_p$
coupling constants from leptonic meson decays. In many cases these bounds are
better than previous bounds. We have summarized the bounds in the tables 
at the end of this text. With the formulae given the bounds can easily 
be updated when the data improve. Furthermore, if additional decays are
 measured (e.g. from the $B$-factories) one can determine additional 
bounds.  Eq.(\ref{result0}) and eq.(\ref{vfr}) can be used to consider 
12 cases: $D^-$ 
($i=1,j=2$), $D^-_s$ ($i=2,j=2$), $B^-$ ($i=3,j=1$), $B^-_c$ ($i=3,j=2$) 
decaying into $e+\nu^C$ and $\mu+\nu^C$ ($f=1,g=2$), $e+\nu^C$ and 
$\tau+\nu^C$ ($f=1,g=3$), $\mu+\nu^C$ and $\tau+\nu^C$ ($f=2,g=3$); 
eq.(\ref{r3}) can be applied 
to the decay of $B^0_s$ ($i=2,j=3$) to $\tau+e^C$ ($f=3,n=1$) or
$\tau+\mu^C$ ($f=3,n=2$), and the decay of the $\Upsilon$, ($i=j=3$)
to $\tau+e^C$ or $\tau+\mu^C$ or $\mu+e^C$ ($f=2,n=1$).

\section{Acknowledgements}

Special thanks to Peter Richardson and Michael Kobel for useful
discussions. M.T. would like to thank the Physikalisches Institut,
Universit\"at Bonn for their kind hospitality during most of this
work, and the Evangelisches Studienwerk and Worcester College,
 University of Oxford, for financial support.
%
%
%
  
$~$\\
\\
\begin{center}
\begin{tabular}{|c|c|c|c|}
\hline
 & & & \\
$ \frac{~\mbox{lower limit}~}{
\big(\mbox{\normalsize{\emph{m}}}_{
\mbox{\tiny{Susy}}}\big/\mbox{100 GeV}\big)^2}$  & $\stackrel{
\mbox{\normalsize{product of}}}{
\mbox{\normalsize{$\not\!\!R_p$ coupling constants}}}$
   &$\frac{~\mbox{upper limit}~}{
\big(\mbox{\normalsize{\emph{m}}}_{
\mbox{\tiny{Susy}}}\big/\mbox{100 GeV}\big)^2}$ & $\stackrel{
\mbox{\normalsize{exchanged}}}{
\mbox{\normalsize{sfermion}}}$  \\
 & & & \\
\hline
$-7.9\times10^{-8}$ &Re$[\lam^{\prime*}_{k11}~\lam_{1k1}$] &
$7.1\times10^{-5}$ &$~~\wtilde{{\ell_{L}^{~k}}} 
\phantom{\Bigg(}$ \\
\hline 
$-7.9\times10^{-5}$ &Re$[\lam^{\prime*}_{311}~\lam_{232}$] &
$-$ &$~~\wtilde{{\ell_{L}^{~k}}} 
\phantom{\Bigg(}$ \\
\hline
0 &$|\lam^{\prime*}_{k11}~\lam_{3k1}|$ &$3.4\times 10^{-6}$ &
$~~\wtilde{{\ell_{L}^{~k}}}\phantom{\Bigg(}$ \\
\hline
0 &$|\lam^{\prime*}_{211}~\lam_{322}|$ &$1.5\times 10^{-3}$ &
$~~\wtilde{{\ell_{L}^{~k}}}\phantom{\Bigg(}$ \\
\hline
0 &$|\lam^{\prime*}_{111}~\lam_{211}|$ &$3.4\times
 10^{-6}$ &$~~\wtilde{{\ell_{L}^{~k}}}\phantom{\Bigg(}$ \\
\hline
$-7.0\times10^{-7}$ &Re$[\lam^{\prime*}_{k12}~\lam_{1k1}$] &
$1.8\times10^{-5}$ &$~~\wtilde{{\ell_{L}^{~k}}} 
\phantom{\Bigg(}$ \\
\hline 
$-1.8\times10^{-4}$ &Re$[\lam^{\prime*}_{k12}~\lam_{2k2}$] &
$3.8\times10^{-3}$, $k=3$ &$~~\wtilde{{\ell_{L}^{~k}}} 
\phantom{\Bigg(}$ \\
\hline
0 &$|\lam^{\prime*}_{k12}~\lam_{2k1}|$ &$5.4\times 10^{-6}$ &
$~~\wtilde{{\ell_{L}^{~k}}}\phantom{\Bigg(}$ \\
\hline
0 &$|\lam^{\prime*}_{k12}~\lam_{3k1}|$ &$5.4\times 10^{-6}$ &
$~~\wtilde{{\ell_{L}^{~k}}}\phantom{\Bigg(}$ \\
\hline
0 &$|\lam^{\prime*}_{k12}~\lam_{1k2}|$ &$1.3\times 10^{-3}$ &
$~~\wtilde{{\ell_{L}^{~k}}}\phantom{\Bigg(}$ \\
\hline
0 &$|\lam^{\prime*}_{k12}~\lam_{3k2}|$ &$1.3\times 10^{-3}$ &
$~~\wtilde{{\ell_{L}^{~k}}}\phantom{\Bigg(}$ \\
\hline
0 &$|\lam^{\prime*}_{313}~\lam_{233}|$ &$2\times10^{-3}$ &
$~~\wtilde{{\ell_{L}^{~k}}}\phantom{\Bigg(}$ \\
\hline
$-6.4\times10^{-4}$ &$\mbox{Re}[\lam^{\prime*}_{213}~\lam_{323}]$ &
$1\times 10^{-3}$ &$~~\wtilde{{\ell_{L}^{~k}}}
\phantom{\Bigg(}$ \\
\hline 
$0$ & $|\lam^{\prime*}_{k13}~\lam_{3k1}|$ & $6\times10^{-4}$&
$~~\wtilde{{\ell_{L}^{~k}}} \phantom{\Bigg(}$ \\
\hline
$0$ & $|\lam^{\prime*}_{k13}~\lam_{3k2}|$ & $7\times10^{-4}$&
$~~\wtilde{{\ell_{L}^{~k}}} \phantom{\Bigg(}$ \\\hline
\end{tabular}
\label{table2}
\end{center}
\newpage\begin{center}
\begin{tabular}{|c|c|c|c|}
\hline
 & & & \\
$ \frac{~\mbox{lower limit}~}{
\big(\mbox{\normalsize{\emph{m}}}_{
\mbox{\tiny{Susy}}}\big/\mbox{100 GeV}\big)^2}$  & $\stackrel{
\mbox{\normalsize{product of}}}{
\mbox{\normalsize{$\not\!\!R_p$ coupling constants}}}$
   &$\frac{~\mbox{upper limit}~}{
\big(\mbox{\normalsize{\emph{m}}}_{
\mbox{\tiny{Susy}}}\big/\mbox{100 GeV}\big)^2}$ & $\stackrel{
\mbox{\normalsize{exchanged}}}{
\mbox{\normalsize{sfermion}}}$  \\
 & & & \\
\hline
0 & $|\lam^\prime_{123}~\lam^{\prime*}_{222}|$ &
$8\times10^{-3}$ &$~~\wtilde{{u_{L}^{~k}}}\phantom{\Bigg(}$ \\
\hline
0 & $|\lam^\prime_{1k2}~\lam^{\prime*}_{2k3}|$,~$k\neq1$ &
$8\times10^{-3}$ &$~~\wtilde{{u_{L}^{~k}}}\phantom{\Bigg(}$ \\
\hline
0 & $|\lam^\prime_{1k2}~\lam^{\prime*}_{2k1}|$ &
$3\times10^{-7}$ &$~~\wtilde{{u_{L}^{~k}}}\phantom{\Bigg(}$ \\
\hline
0 & $|\lam^\prime_{1k1}~\lam^{\prime*}_{2k2}|$ &
$3\times10^{-7}$ &$~~\wtilde{{u_{L}^{~k}}}\phantom{\Bigg(}$ \\
\hline\hline
0 &$|\lam^{\prime*}_{k32}~\lam_{k12}|$ &$7\times10^{-5}$ &
$~~\wtilde{{\nu_{L}^{~k}}}\phantom{\Bigg(}$ \\
\hline
0 &$|\lam^{\prime*}_{k23}~\lam_{k21}|$ &$7\times10^{-5}$ &
$~~\wtilde{{\nu_{L}^{~k}}}\phantom{\Bigg(}$ \\
\hline
0 &$|\lam^{\prime*}_{k32}~\lam_{k21}|$ &$7\times10^{-5}$ &
$~~\wtilde{{\nu_{L}^{~k}}}\phantom{\Bigg(}$ \\
\hline
0 &$|\lam^{\prime*}_{k23}~\lam_{k12}|$ &$7\times10^{-5}$ &
$~~\wtilde{{\nu_{L}^{~k}}}\phantom{\Bigg(}$ \\
\hline
0 &$|\lam^{\prime*}_{k21}~\lam_{k12}|$ &$6\times10^{-9}$ &
$~~\wtilde{{\nu_{L}^{~k}}}\phantom{\Bigg(}$ \\
\hline
0 &$|\lam^{\prime*}_{k12}~\lam_{k12}|$ &$6\times10^{-9}$ &
$~~\wtilde{{\nu_{L}^{~k}}}\phantom{\Bigg(}$ \\
\hline
0 &$|\lam^{\prime}_{k12}~\lam_{k21}^*|$ &$6\times10^{-9}$ &
$~~\wtilde{{\nu_{L}^{~k}}}\phantom{\Bigg(}$ \\
\hline
0 &$|\lam^{\prime}_{k21}~\lam_{k21}^*|$ &$6\times10^{-9}$ &
$~~\wtilde{{\nu_{L}^{~k}}}\phantom{\Bigg(}$ \\
\hline
0 &$|\lam^{\prime*}_{311}~\lam_{312}|$ &$3\times10^{-3}$ &
$~~\wtilde{{\nu_{L}^{~k}}}\phantom{\Bigg(}$ \\
\hline
0 &$|\lam^{\prime}_{311}~\lam_{321}^*|$ &
$3\times10^{-3}$ &$~~\wtilde{{\nu_{L}^{~k}}}
\phantom{\Bigg(}$ \\ \hline
\end{tabular}
\label{table1}
\end{center}

\end{document}